# FABRICATION OF ZINC OXIDE/GRAPHENE-CARBON NANOTUBES NANOCOMPOSITE WITH ENHANCED DYE DEGRADATION ABILITY

**Intekhab Alam**[1], **Md. Minaruzzaman**[1], **Md. Ashiqur Rahman**[1], **and M. A. Basith**[2]

[1]Department of Mechanical Engineering (ME), Bangladesh University of Engineering and Technology (BUET), Dhaka-1000, Bangladesh
[2]Nanotechnology Research Laboratory, Department of Physics, Bangladesh University of Engineering and Technology (BUET), Dhaka-1000, Bangladesh
intekhabsanglap@gmail.com*, mdminaruzzaman1@gmail.com, ashiqurrahman@me.buet.ac.bd, and mabasith@phy.buet.ac.bd

***Abstract***- *A comparative study between zinc oxide/graphene-carbon nanotubes (ZnO/Gr-CNTs) nanocomposite and ZnO nanoparticles was carried out to investigate their abilities in the degradation of Rhodamine B (RhB) dye. The modified Hummer's method was utilized to prepare the graphene oxide (GO) nanosheet. Moreover, Gr-CNTs had been synthesized from GO and multi-walled carbon nanotubes with hydroxyl group (MWCNTs-OH). The hydrothermal method was employed to fabricate the ZnO/Gr-CNTs nanocomposite from ZnO nanoparticles and Gr-CNTs. During the characterization by X-ray Diffraction (XRD), all the significant peaks of ZnO and ZnO/Gr-CNTs were found in the same phase angle. In addition, the final nanocomposite was also characterized by Field Emission Scanning Electron Microscope (FESEM) and Energy Dispersive X-ray Spectroscopy (EDS). Finally, from the dye degradation test, it was apparent that ZnO/Gr-CNTs nanocomposite exhibited superior dye removal ability compared to the ZnO nanoparticles.*

## 1. INTRODUCTION

Synthetic dyes are widely used by the textile industry as they are inexpensive, brighter, and easier to apply. However, these synthetic dyes have good solubility in water that makes them common water pollutants [1]. Their existence in the industrial wastewaters causes serious environmental hazards because of their mutagenicity to humans and toxicity to aquatic life [2]. Therefore, it is mandatory to remove dyes from water prior to their discharge into water bodies. Adsorption, photocatalytic degradation, biodegradation, advanced oxidation, electrochemical, and chemical oxidation are generally utilized processes for dye removal [3]. Among them, the photocatalysis process has attracted researchers because of its cost-effectiveness and biodegradable and non-toxic end products. ZnO, $TiO_2$, $SrO_2$, $WO_3$, $Fe_2O_3$, $ZrO_2$, CdS, $SrTiO_3$, and ZnS have been employed as heterogeneous photocatalysts for the removal of several variants of pollutants and dyes from wastewater [1], [4]. Environmentally benign ZnO has a wide band-gap (~3.4 eV), good stability, low cost, and excellent optoelectronic properties [5]. All of these properties make ZnO a potential candidate for the photocatalysis and degradation of various dyes. However, ZnO is not an efficient dye degradant on its own because of its rapid recombination of electron-hole pairs, insufficient absorption efficiency in the visible sunlight region, and other chemical and physical obstacles [3].

Meanwhile, adsorption is one of the most efficient and cost-effective processes for the degradation of textile dyes. For the adsorption of different dyes, carbon-based inorganic supports have been widely employed [2]. Among the carbonaceous materials utilized as adsorbents, the relatively new graphene oxide (GO) and carbon nanotubes (CNTs) have attained much attention as they have superior adsorption capabilities due to their large specific surface area and the presence of a wide spectrum of surface functional groups [6]. Moreover, the adsorption of textile dye on GO and multiwalled carbon nanotubes (MWCNTs) have been ascribed to $\pi-\pi$ electron donor-acceptor interactions and electrostatic interaction [7]. It has been reported that GO can absorb the CNTs by the strong $\pi-\pi$ interaction to construct graphene-carbon nanotubes (Gr-CNTs) hybrid structure [8]. The hybridized structure has shown superior performance than its constituents CNTs and GO and can be applied as transparent conductive electrodes for various applications [9]. Previously, Chen *et al.* have reported the preparation of Gr/MWCNTs/ZnO nanocrystalline aggregate by the spray-drying method. The Gr/MWCNTs hybrid drastically enhanced the adsorption capacity and photocatalytic activity of ZnO nanoparticles and made it an excellent degradant of methyl orange dye [8]. However, in this study, ZnO/Gr-CNTs nanocomposite was fabricated by simple hydrothermal method for the first time, considering the potentials of ZnO nanoparticles and Gr-CNTs hybrid [10]. Different characterization tests were performed on the final nanocomposite, and its efficiency to degrade Rhodamine B (RhB) dye was explored to find its suitability in wastewater treatment.

## 2. EXPERIMENTAL PROCEDURE

### 2.1 Synthesis of GO using modified Hummer's method

GO was synthesized by modified Hummer's method that was previously reported by Paulchamy *et al.* [11]. This method of synthesis consists of both oxidation and exfoliation of graphite sheets because of the thermal treatment of the solution. For fabricating GO by this method, 2 g graphite nanoparticles (99.9% purity, US Research Nanomaterials) and 2 g $NaNO_3$ were mixed in 90 mL of $H_2SO_4$ (98% concentration) in a 1000 mL volumetric flask initially. Then the volumetric flask with the mixture was kept in an ice bath under 5 °C temperature with continuous stirring with a magnetic stirrer for 4 hours. Subsequently, 12 g $KMnO_4$ was mixed with the suspension at a slow rate, which was regulated cautiously for maintaining the reaction temperature under 15 °C. After that, 184 mL deionized (DI) water was added to the mixture very slowly for diluting the mixture that was maintained under stirring for 2 hours. Next, the mixture was stirred at 35 °C temperature for 2 hours after removing the ice bath, and it was kept in a reflux system at 98 °C for 10 minutes afterward. The temperature was altered to 30 °C after 10 minutes to get a brown-colored solution. Again, the temperature was modulated to 25 °C after 10 minutes, which was then kept for 2 hours. Afterward, the solution was treated with 40 mL $H_2O_2$ that altered the color to bright yellow. An equal amount of prepared solution was added to 200 mL of DI water taken in two separate beakers, which were stirred for 1 hour. Then the solution mixed with water was kept unstirred for 3-4 hours for settling the particles at the bottom. Subsequently, the separated water was poured into another beaker for filtering purposes. Next, for washing the mixture, centrifugation with HCl (10% concentration) was performed repeatedly. After that, the mixture was also washed by centrifugation with DI water multiple times. Finally, a pH-neutral gel-like substance was formed, which was then placed in a Teflon-lined autoclave and vacuum-dried at 60 °C temperature using a programmable oven for 6 hours to obtain the GO powder.

### 2.2 Fabrication of Gr-CNTs and ZnO/Gr-CNTs

For the fabrication of Gr-CNTs, primarily, 0.1 g of functionalized MWCNTs with 5.58 wt% hydroxyl groups (MWCNTs-OH) (95% purity, US Research Nanomaterials) was exfoliated in 40 g of N, N-dimethylformamide (DMF) via ultrasonic treatment. Moreover, 0.1 g of GO, synthesized from graphite nanoparticles utilizing the Modified Hummers method, was also exfoliated into the GO nanoparticles in 10 g of DMF with the assistance of ultrasonic treatment. The mixed solution (2.5 mg/g) consisting of MWCNTs-OH and GO nanoparticles had been placed in a Teflon-lined autoclave and dried for 24 hours at 100 °C using the oven to finally fabricate the Gr-CNTs as shown in Figure 1(a) [10], [12]. As the oxygenic functional groups on the GO sheet are removed under heat treatment, the GO is reduced to graphene (Gr) in Gr-CNTs [13]. Next, ZnO/Gr-CNTs nanocomposite was synthesized using the Hydrothermal method by mixing ZnO nanoparticles (99.9% purity, 35-45 nm average particle size, US Research Nanomaterials) and previously fabricated Gr-CNTs in the mass ratios of 5:1. After that, the mixture was added to 60 mL DI water, which was sealed in a 100 mL Teflon vessel. Then it was placed in the autoclave and heated by the oven at 160 °C for 18 hrs. Subsequently, the resultant product was collected via centrifugation and washed with DI water. After freeze-drying the product for 24 hours, ZnO/Gr-CNTs nanocomposite was finally synthesized as shown in Figure 1(b) [10].

### 2.3 Characterizations

The structural properties of ZnO nanoparticles and ZnO/Gr-CNTs nanocomposite were observed by analyzing their powder X-Ray Diffraction (XRD) patterns. These were achieved by utilizing a diffractometer (PANalytical Empyrean) with a Cu X-ray source. Furthermore, the field emission scanning electron microscopy (FESEM) imaging of the ZnO/Gr-CNTs nanocomposite was carried out with the assistance of a scanning electron microscope (JSM 7600F, JEOL-Japan). For the elemental characterization of ZnO/Gr-CNTs nanocomposite, energy-dispersive X-ray spectroscopy (EDS) was performed by an X-ray spectroscope attached to the aforementioned scanning electron microscope.

### 2.4 Dye degradation experiment

For the analysis of degradation performance of RhB dye utilizing the synthesized ZnO nanoparticles, at first 10 mg RhB was dissolved in 50ml of DI water [14], [15]. Then 5 mL of the solution was extracted for measuring its absorbance with the help of a UV-vis spectrophotometer (UV-2600, Shimadzu). The intensity of the absorbance peak (553nm) was proportional to the amount of RhB present in the solution, and the same procedure was utilized each time to find out the remaining amount of RhB in the solution. Subsequently, for obtaining a homogenous solution, the solution was magnetically stirred in dark conditions for 30 minutes. Then 25 mg of the ZnO nanoparticles photocatalyst with 1 mL of 0.5 M HCl was mixed with the solution, which was further stirred for 1 hour for confirming an adsorption-desorption equilibrium. After that, a mercury-xenon lamp (Hamamatsu L8288, 500 W) with a fixed irradiance value of 100 mW/cm$^2$ was employed as the solar simulator to irradiate the solution for commencing the photocatalysis process.

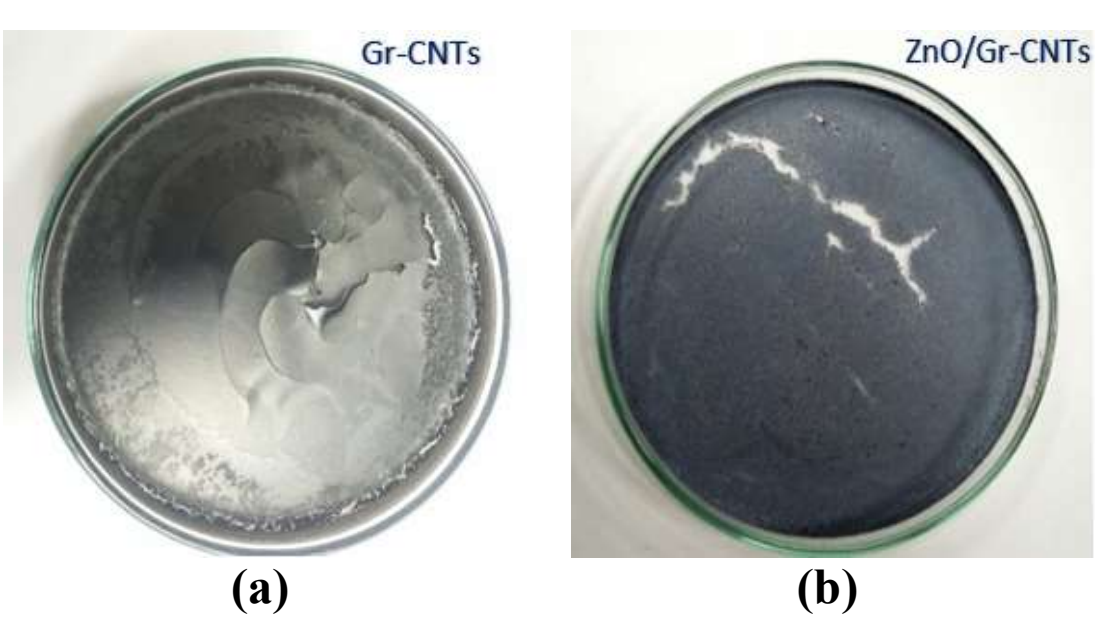

Fig. 1: Images illustrating the fabricated (a) Gr-CNTs

and (b) ZnO/Gr-CNTs

The nanomaterials were precipitated by a centrifuge machine at 5500 rpm for 10 minutes before the absorbance measurement. Afterward, the absorbance measurement of the solution was performed at every 1-hour interval to observe the degradation of RhB. Finally, the remaining photocatalyst nanomaterials in the suspension were separated by centrifugation, washed with DI water, and dried later before recycling the photocatalyst. However, there were some differences while performing the dye degradation test of ZnO/Gr-CNTs nanocomposite. By following the previously explained procedure, 25 mg of the ZnO/Gr-CNTs nanocomposite with 1 mL of 0.5 M HCl was added to a homogenous RhB solution. But after stirring the solution in the dark medium, it was observed that the percentage of degradation was very high in dark conditions. Therefore, the degradation of RhB molecules by adsorption to the ZnO/Gr-CNTs nanocomposite was highly significant, and so, the rest of the experiment was run in the dark condition without the presence of a mercury-xenon lamp [7], [14]. The other procedures were the same, except that the absorbance measurement of the solution was carried out by the UV-vis spectrophotometer at every 15 minutes interval instead of 1 hour for monitoring the rapid degradation of RhB in the dark medium.

## 3. RESULTS AND DISCUSSION

### 3.1 Structural characterizations

The XRD patterns of ZnO/Gr-CNTs nanocomposite and ZnO nanoparticles have been illustrated in Figures 2(a) and 2(b), respectively. The X-ray photons incident on the crystal observe the lattice as a diffraction grating and produce a diffraction pattern with peaks at specific Bragg angles because a crystal lattice is a periodic arrangement of the constituent ions [15]. It was observed from the XRD patterns that both of the samples showed the diffraction peaks of ZnO corresponding to (100), (002), (101), (102), (110), (103), (200), (112), and (201) planes, respectively. By comparing with the standard data provided by JCPDS 36-1451, all diffraction peaks of the samples were in good agreement with those of the hexagonal wurtzite structure of ZnO [8]. The most intense peak was visible at about $2\theta = 36.24^0$ at (101) plane, and other significant peaks were observed at about 31.77°, 34.42°, and 56.62° corresponding to (100), (002), and (110) planes, respectively. There were significant similarities between the XRD patterns of ZnO/Gr-CNTs nanocomposite and ZnO nanoparticles according to Figure 2. The positions of the characteristic peaks of ZnO/Gr-CNTs nanocomposite were in situ with that of ZnO nanoparticles, which indicates that the presence of Gr-CNTs did not alter the crystal phase of ZnO [3]. However, the intensity of the peaks of ZnO nanoparticles reduced after the incorporation of Gr-CNTs in the final nanocomposite.

### 3.2 Morphological analysis

The FESEM imaging of the ZnO/Gr-CNTs nanocomposite has been shown in Figure 3(a) for determining the morphology of the final nanocomposite. It could be observed that the ZnO/Gr-CNTs nanocomposite had a grain-like nanocrystalline structure where agglomerations were present. ImageJ software was utilized to analyze the particle size from the FESEM image. A histogram was drawn from the image as shown in Figure 3(b) where most of the particles were within the 20 nm to 50 nm range. The calculated average particle size was approximately 42 nm. Meanwhile, the average particle size of ZnO nanoparticles was from 35 nm to 45 nm (US Research Nanomaterials). Therefore, there was no significant alteration in the average particle size of the ZnO/Gr-CNTs nanocomposite from the ZnO nanoparticles, which was previously suggested by the XRD results. In addition, the XRD and FESEM confirmed the optimal presence of Gr-CNTs as the crystalline quality of ZnO in the final nanocomposite was unperturbed [3].

The EDS spectra of the final nanocomposite have been illustrated in Figure 4, whereas the elemental characterization results obtained from the EDS have been shown in Table 1. The EDS results implied the presence of oxygen, zinc, and carbon in the nanocomposite. Also, the highest percentage of carbon among other materials might be due to the utilization of carbon strips during the EDS measurement. Another reason could be that the measurement was taken from the surface location of the final nanocomposite where an excessive amount of carbon from Gr-CNTs was present.

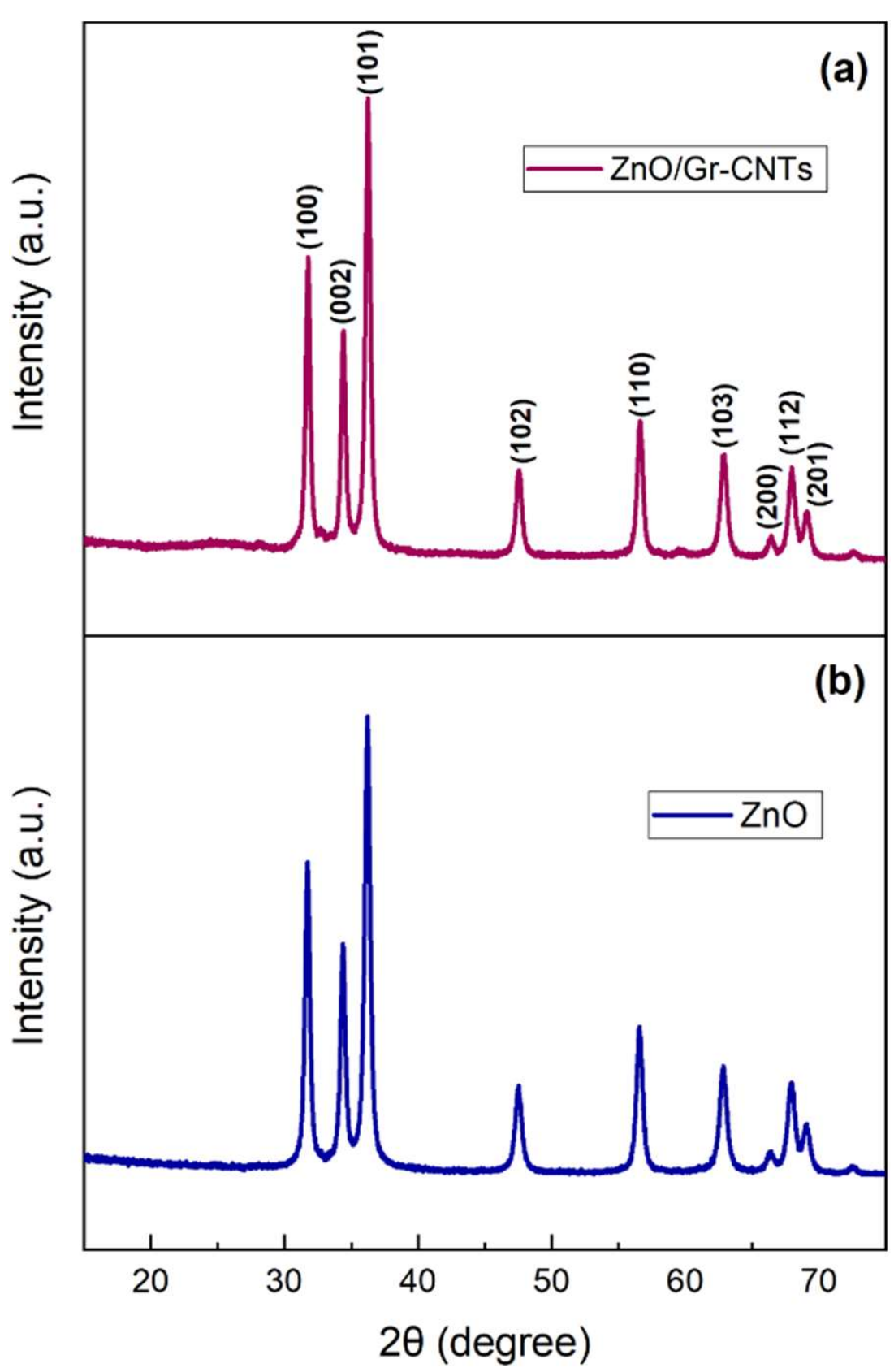

Fig. 2: XRD patterns of the fabricated (a) ZnO/Gr-CNTs

nanocomposite and (b) ZnO nanoparticles

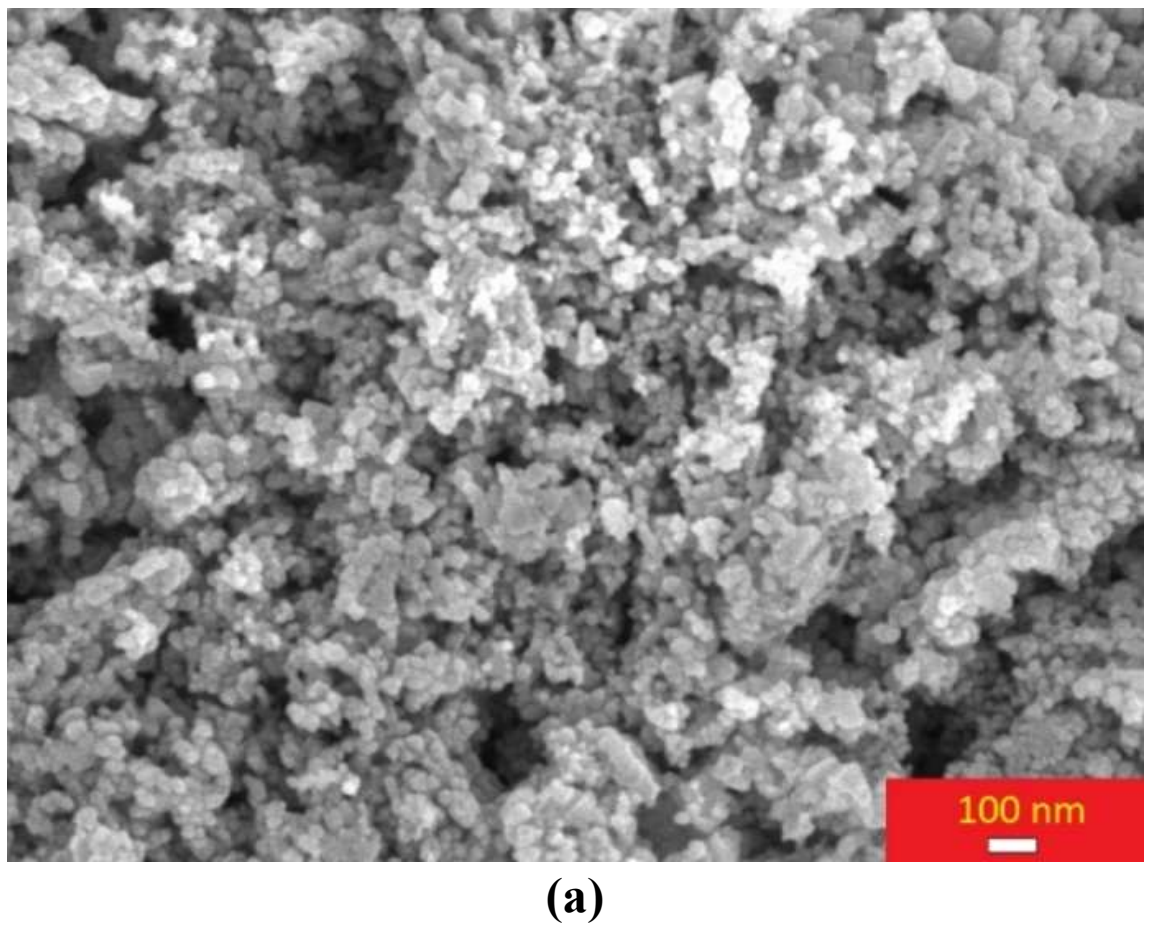

**(a)**

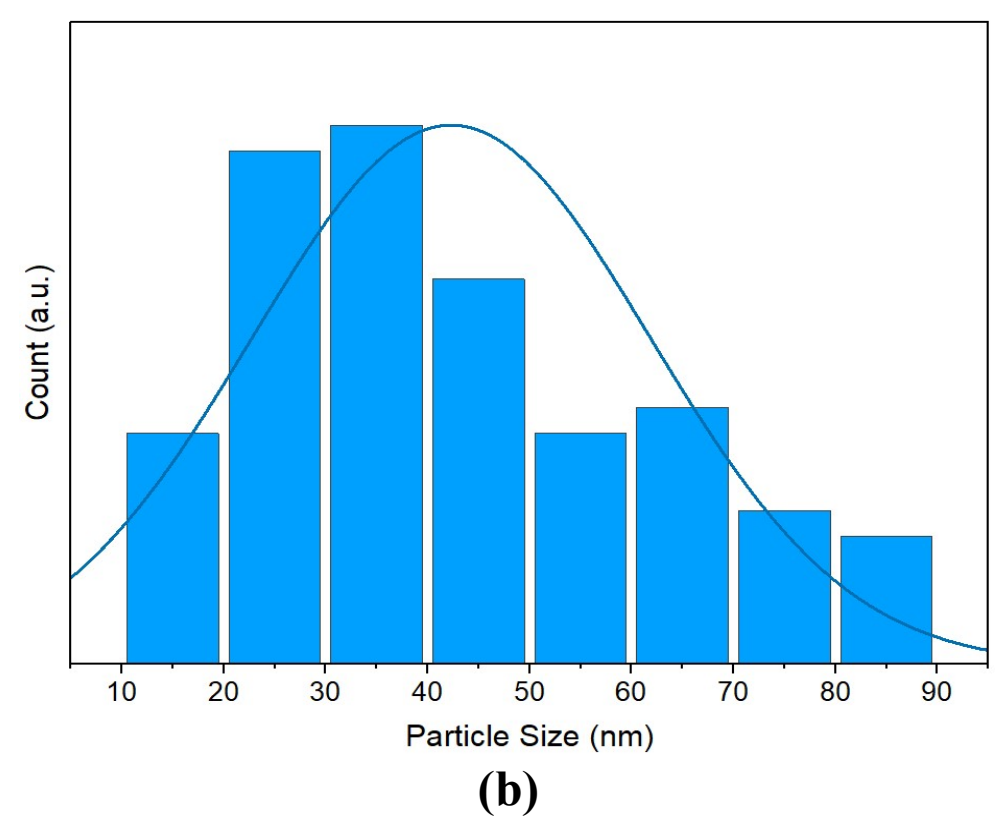

**(b)**

Fig. 3: (a) FESEM image illustrating the surface morphology of ZnO/Gr-CNTs and (b) corresponding histogram utilized for calculating the particle size

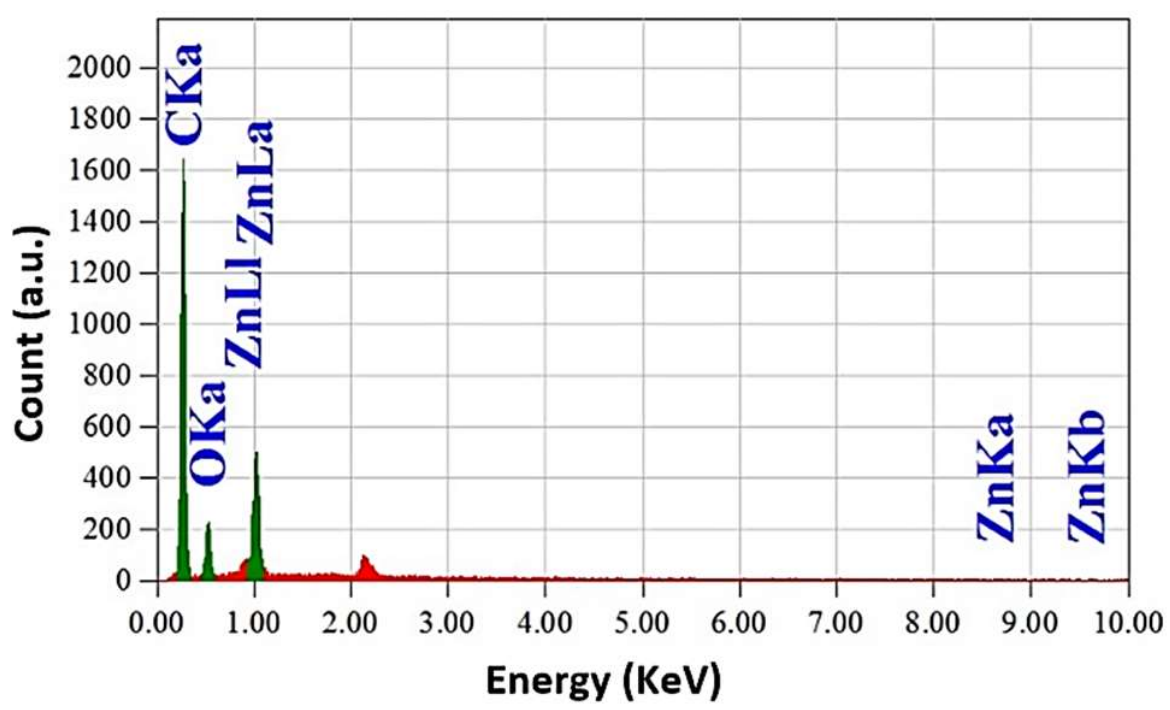

Fig. 4: The EDS spectra of the fabricated ZnO/Gr-CNTs

Table 1: The elemental characterization results of ZnO/Gr-CNTs nanocomposite obtained from EDS

| Element | Energy (KeV) | Mass (%) | Atom (%) |
| --- | --- | --- | --- |
| C K | 0.277 | 71.47 | 85.07 |
| O K | 0.525 | 12.87 | 11.50 |
| Zn L | 1.012 | 15.66 | 3.42 |

### 3.3 Dye degradation

The RhB dye was degraded with the help of ZnO nanoparticles and ZnO/Gr-CNTs nanocomposite. Figure 5(a) illustrates the absorption spectra for RhB solution in the presence of ZnO nanoparticles in the dark and light mediums. It was observed that there was no significant fluctuation in the absorption peak intensity after 1 hour irradiation from the 1 hour in the dark medium. However, after 2 hours of irradiation, the absorption peak intensity declined sharply. Then the absorption peak intensity faced a gradual reduction with the increasing exposure time. Contrarily, Figure 5(b) shows the absorption spectra for RhB solution in the dark medium in the presence of ZnO/Gr-CNTs nanocomposite. It was apparent that the absorption peak intensity sharply reduced after 15 minutes, and afterward, it declined gradually. Furthermore, the maximum intensity ratio, $C/C_0$ with varying time was plotted to determine the dye degradation efficiency. Here, C and $C_0$ are the concentrations of RhB at the time t = t and t = 0, respectively. Also, the $C/C_0$ was calculated considering the absorbance maxima. In Figures 6(a) and 6(b), the $C/C_0$ with varying time for the degradation reactions in the presence of ZnO nanoparticles and ZnO/Gr-CNTs nanocomposite have been illustrated.

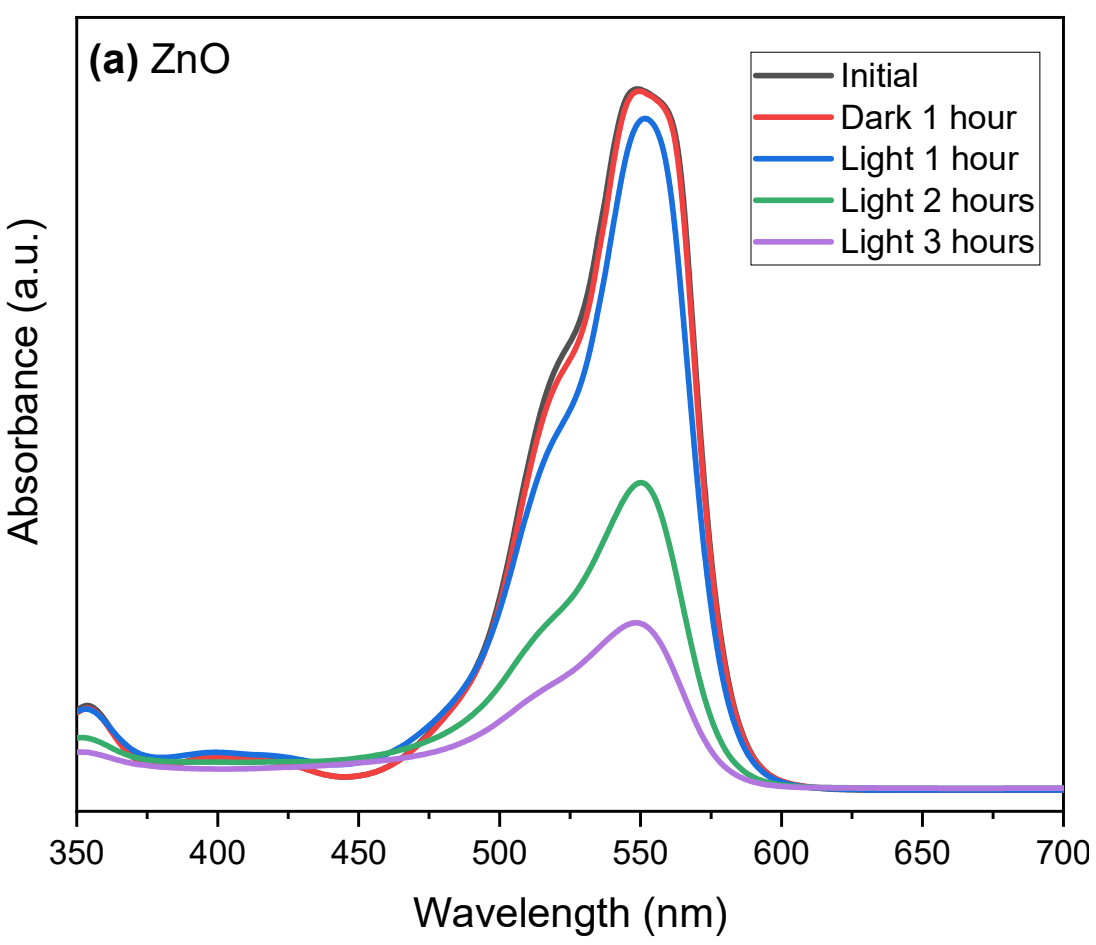

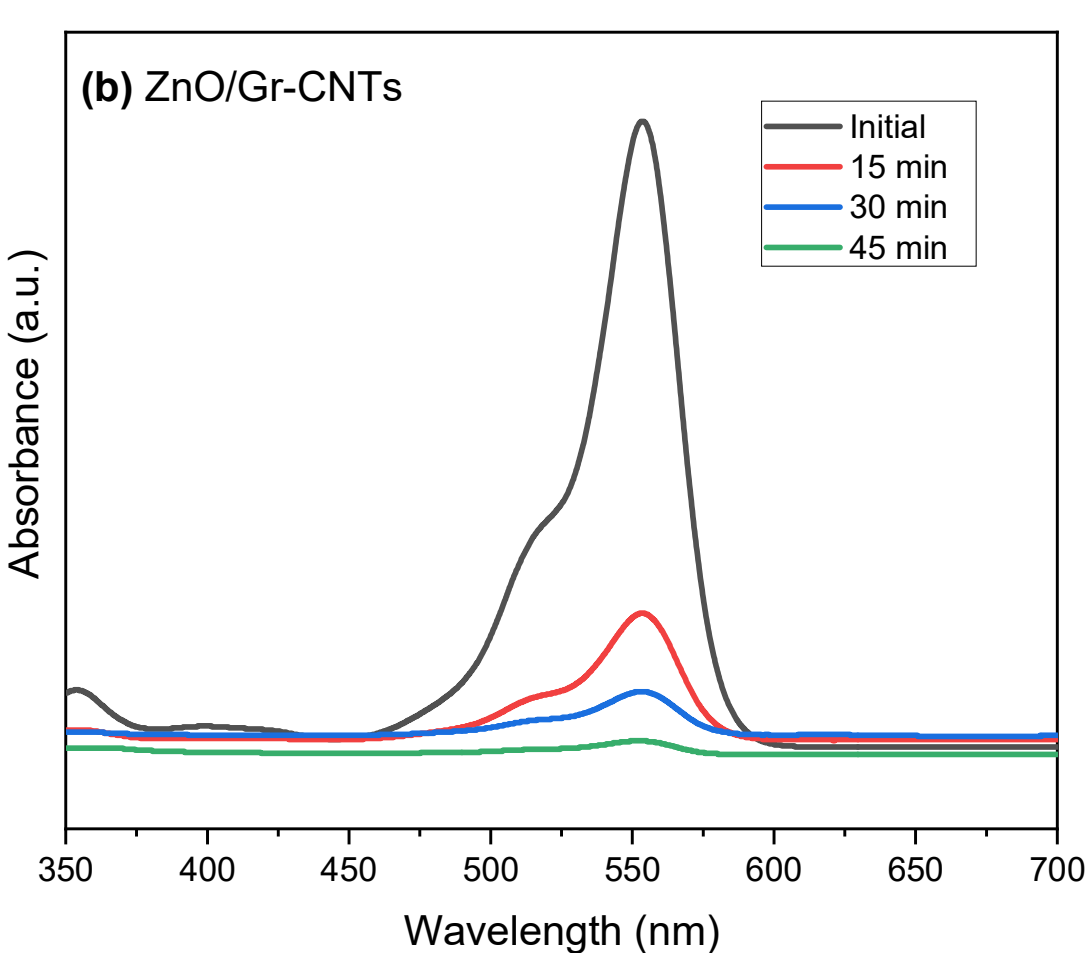

Fig. 5: The UV-Vis absorbance spectra for RhB solutions with (a) ZnO and (b) ZnO/Gr-CNTs at

different time intervals

It was observed that ZnO nanoparticles degraded 77% of the RhB dye initially present in the solution after 3 hours of illumination, while ZnO/Gr-CNTs nanocomposite was able to degrade 95% of the RhB dye after only 45 minutes in the dark medium. The degradation due to the adsorption of RhB molecules by ZnO nanoparticles was very marginal as evident from the dark test. Therefore, the decolorization of RhB dye by illuminated ZnO nanoparticles could be significantly attributed to photocatalytic degradation. The photodegradation of RhB dye is caused by certain redox reactions that could be catalyzed in the presence of ZnO nanoparticles [15]. Conversely, the rapid degradation of RhB dye in the dark medium by ZnO/Gr-CNTs nanocomposite indicated that the degradation of the dye via adsorption to the final nanocomposite was highly significant. The difference in dye removal ability could be attributed to the inclusion of Gr-CNTs as both GO and MWCNTs are well-known strong adsorbents [7]. Moreover, the presence of Gr-CNTs could increase the specific surface area of ZnO nanoparticles significantly. The increasing specific surface area and superior adsorption capacity could be attributed to the incorporation of MWCNTs onto Gr sheets by $\pi$–$\pi$ interaction, thus preventing the restacking of Gr sheets and improving their surface area [8].

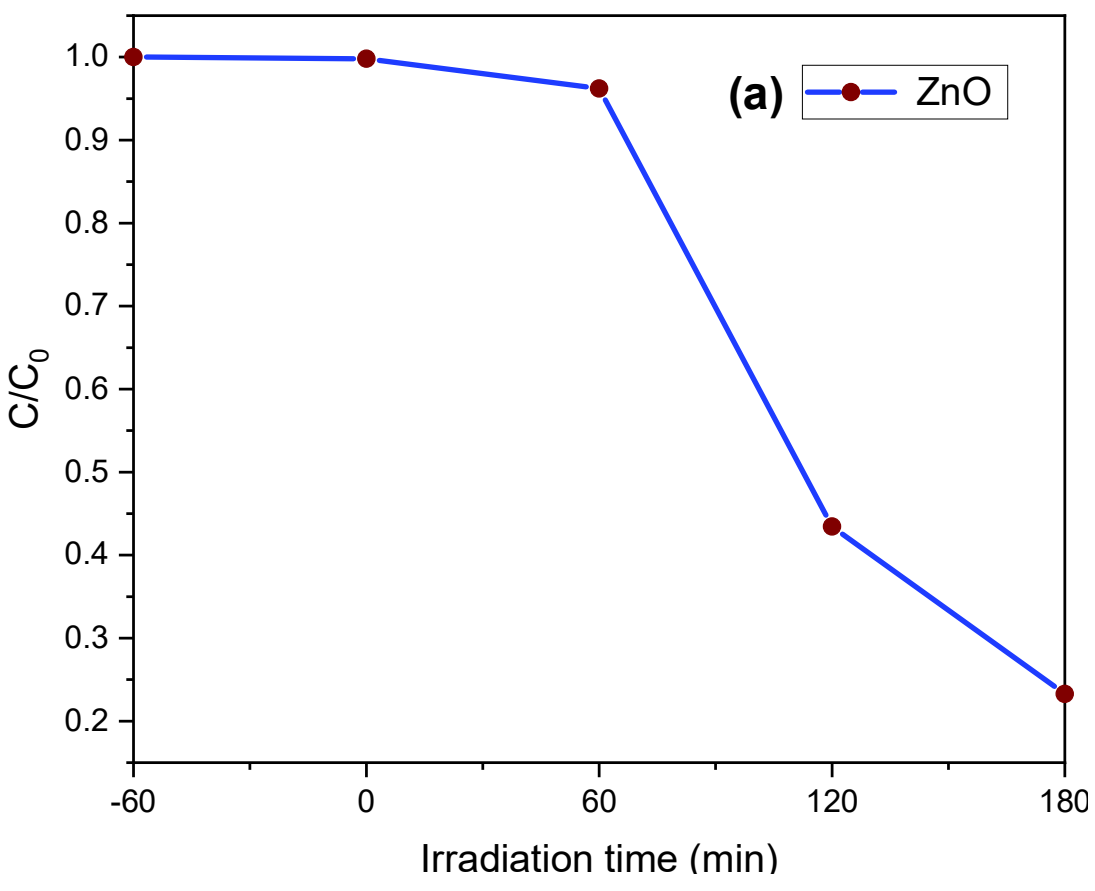

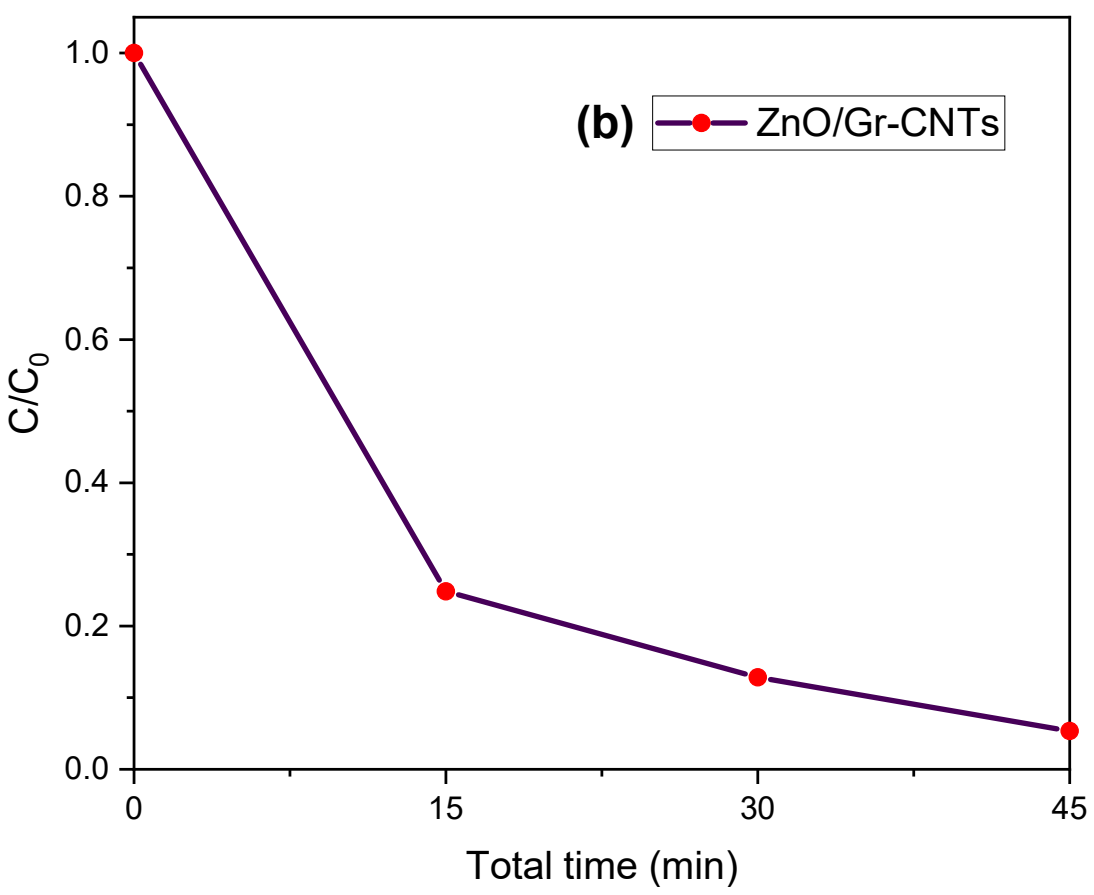

Fig. 6: The degradation of RhB by (a) ZnO and (b) ZnO/Gr-CNTs with varying time

## 4. CONCLUSIONS

In this study, GO nanosheet, Gr-CNTs, and finally, ZnO/Gr-CNTs nanocomposite were experimentally fabricated. The XRD patterns of ZnO/Gr-CNTs nanocomposite and ZnO nanoparticles were similar to each other, suggesting that the crystal phase of ZnO was unperturbed in the presence of Gr-CNTs. From the FESEM image, a grain-like nanocrystalline structure containing agglomerations was visible for the final nanocomposite. Moreover, the average particle size of the final nanocomposite was about 42 nm. According to EDS, the final nanocomposite contained Zn, C, and O, among which C had the highest percentage. During the dye degradation test, ZnO/Gr-CNTs nanocomposite showed strong adsorption ability and degraded 95% of the RhB dye after only 45 minutes in the dark medium. In comparison, ZnO nanoparticles exhibited photocatalytic activity and required 3 hours of illumination to photodegrade only 77% of the RhB dye. This implied that, the degradation of RhB via ZnO nanoparticles was photocatalytic and that by ZnO/Gr-CNTs was due to adsorption. Therefore, the final nanocomposite showed impressive dye degradation ability that can be applied for wastewater treatment, sewage treatment, and environmental protection without the presence of any illumination.

## 5. ACKOWLEDGEMENT

The authors sincerely acknowledge Bangladesh University of Engineering and Technology (BUET) and Nanotechnology Research Laboratory (NRL) for facilitating this work.

## 6. REFERENCES

[1] S. Rani, M. Aggarwal, M. Kumar, S. Sharma, and D. Kumar, “Removal of methylene blue and rhodamine B from water by zirconium oxide/graphene,” *Water Science*, vol. 30, no. 1, pp. 51–60, Apr. 2016, doi: 10.1016/j.wsj.2016.04.001.

[2] K. Singh and S. Arora, “Removal of synthetic textile dyes from wastewaters: A critical review on present treatment technologies,” *Critical Reviews in Environmental Science and Technology*, vol. 41, no. 9, Apr. 2011, doi: 10.1080/10643380903218376.

[3] N. J. Salim, I. Zarin, M. A. Rahman, and M. A. Basith, “Photocatalytic treatment of textile effluent using reduced graphene oxide and zinc oxide integrated nanocomposite with inherent semiconductivity,” in *AIP Conference Proceedings*, Jul. 2018, vol. 1980. doi: 10.1063/1.5044295.

[4] I. Alam and M. A. Ashraf, “Effect of different device parameters on tin-based perovskite solar cell coupled with In2S3 electron transport layer and CuSCN and Spiro-OMeTAD alternative hole transport layers for high-efficiency performance,” *Energy Sources, Part A: Recovery, Utilization and Environmental Effects*, 2020, doi: 10.1080/15567036.2020.1820628.

[5] I. Alam, R. Mollick, and M. A. Ashraf, “Numerical simulation of Cs2AgBiBr6-based perovskite solar cell with ZnO nanorod and P3HT as the charge transport

layers," *Physica B: Condensed Matter*, vol. 618, 2021, doi: 10.1016/j.physb.2021.413187.
[6] K. Pyrzynska, "Sorption of Cd(II) onto carbon-based materials-a comparative study," *Microchimica Acta*, vol. 169, no. 1–2, Apr. 2010, doi: 10.1007/s00604-010-0305-5.
[7] X. Ren, J. Li, X. Tan, and X. Wang, "Comparative study of graphene oxide, activated carbon and carbon nanotubes as adsorbents for copper decontamination," *Dalton Transactions*, vol. 42, no. 15, pp. 5266–5274, Apr. 2013, doi: 10.1039/c3dt32969k.
[8] C. S. Chen *et al.*, "Synthesis and photocatalytic property of graphene/multi-walled carbon nanotube/ZnO nanocrystalline aggregates hybrids by spray drying method," *Functional Materials Letters*, vol. 7, no. 4, 2014, doi: 10.1142/S1793604714500489.
[9] H. X. Kong, "Hybrids of carbon nanotubes and graphene/graphene oxide," *Current Opinion in Solid State and Materials Science*, vol. 17, no. 1, pp. 31–37, Feb. 2013, doi: 10.1016/j.cossms.2012.12.002.
[10] Z. Fu, H. Wang, Y. Wang, S. Wang, Z. Li, and Q. Sun, "Construction of three-dimensional g-C3N4/Gr-CNTs/TiO2 Z-scheme catalyst with enhanced photocatalytic activity," *Applied Surface Science*, vol. 510, Apr. 2020, doi: 10.1016/j.apsusc.2020.145494.
[11] P. B. Arthi G and L. BD, "A Simple Approach to Stepwise Synthesis of Graphene Oxide Nanomaterial," *Journal of Nanomedicine & Nanotechnology*, vol. 06, no. 01, 2015, doi: 10.4172/2157-7439.1000253.
[12] D. Cai, M. Song, and C. Xu, "Highly conductive carbon-nanotube/graphite-oxide hybrid films," *Advanced Materials*, vol. 20, no. 9, pp. 1706–1709, May 2008, doi: 10.1002/adma.200702602.
[13] Y. Zhang *et al.*, "Microwave-assisted, environmentally friendly, one-pot preparation of Pd nanoparticles/graphene nanocomposites and their application in electrocatalytic oxidation of methanol," *Catalysis Science & Technology*, vol. 1, no. 9, 2011, doi: 10.1039/c1cy00296a.
[14] A. M. Tama, S. Das, S. Dutta, M. D. I. Bhuyan, M. N. Islam, and M. A. Basith, "MoS2 nanosheet incorporated α-Fe2O3/ZnO nanocomposite with enhanced photocatalytic dye degradation and hydrogen production ability," *RSC Advances*, vol. 9, no. 69, pp. 40357–40367, 2019, doi: 10.1039/c9ra07526g.
[15] R. Ahsan, A. Mitra, S. Omar, M. Z. Rahman Khan, and M. A. Basith, "Sol-gel synthesis of DyCrO3 and 10% Fe-doped DyCrO3 nanoparticles with enhanced photocatalytic hydrogen production abilities," *RSC Advances*, vol. 8, no. 26, pp. 14258–14267, 2018, doi: 10.1039/c8ra01585f.